\documentclass[aps,prl,twocolumn,floatfix,superscriptaddress]{revtex4}
\usepackage{graphicx}
\usepackage{epsfig}
\begin{document}
\newcommand{\norm}[1]{\ensuremath{| #1 |}}
\title{Spin-charge separation in cold Fermi-gases: a real time analysis}

\author{C. Kollath}
\affiliation{Institute for Theoretical Physics C, RWTH Aachen,
D-52056 Aachen, Germany}
\author{U. Schollw\"ock}
\affiliation{Institute for Theoretical Physics C, RWTH Aachen,
D-52056 Aachen, Germany}
\author{W. Zwerger}
\affiliation{Physics Department, Technical University Munich, D-85748 Garching, Germany}
\date{\today}

\begin{abstract}
Using the adaptive time-dependent density-matrix renormalization
group method for the 1D Hubbard model, the splitting of local
perturbations into separate wave packets carrying charge and spin
is observed in real-time.  We show the robustness of this
separation beyond the low-energy Luttinger liquid theory by
studying the time-evolution of single particle excitations and
density wave packets. %%% of finite strength and at length scales down to a few lattice spacings.
A striking signature of spin-charge separation is found in 1D cold
Fermi gases in a harmonic trap at the boundary between liquid and
Mott-insulating phases. %%%, where the charge is perfectly reflected.
We give quantitative estimates for an experimental observation of
spin-charge separation in an array of atomic wires.
\end{abstract}

\maketitle
One-dimensional (1D) quantum many-body systems have been at the center of
theoretical and experimental interest for the last two decades. Following the
seminal work of Haldane \cite{Hald81}, it has been understood that -
independent of their bosonic or fermionic nature - the low-energy behaviour
of 1D quantum liquids is universally described by the so-called
Luttinger liquid (LL) picture \cite{Voit95,Giam04}.
Probably the most remarkable prediction is the phenomenon of spin-charge separation
in the case of Fermions, i.e. the fact that  - at low energy - the excitations of charge
and spin completely decouple and propagate with different velocities.
%This is in striking contrast to Fermi liquids, where the
%elementary quasiparticle excitations carry both charge and spin,
%moving at the Fermi velocity.
A definite signature of spin-charge separation requires the
observation of the two corresponding branches of excitations in
the single particle spectral function \cite{Zacher}. In condensed matter
systems, numerous experiments have looked  for spin-charge
separation e.g. via photoemission from  1D metallic wires on
surfaces \cite{Sego99}, in 1D organic wires \cite{Lore02}, in
carbon-nanotubes \cite{Bock99}, and in quantum wires in
semiconductors, where the singular nature of the spectral
functions associated with spin-charge-separation was observed in
tunneling experiments \cite{Ausl05}. In the last few years,
ultracold gases in optical lattices are providing an entirely new
area of physics where strong correlations can be studied with
unprecedented control and tunability of the parameters. In
particular these systems open the possibility to investigate the
transition between three, quasi-two and quasi-one dimension.
Recently, an 'atomic quantum wire' configuration in an array of
thousands of parallel atom waveguides was realized in ultracold
Fermi gases by the application of a strong two dimensional optical
lattice \cite{Mori05}. The possibility to use cold atoms for
studying the phenomenon of spin-charge separation was first
suggested by Recati et al.\ \cite{Reca03}. Their analysis is
essentially based on the hydrodynamic Hamiltonian of the LL; the
inhomogeneity due to the presence of a harmonic trap is treated
within a local density approximation (see also \cite{Keck04}). In
practice, with typically less than 100 atoms per atomic wire
\cite{Mori05}, observable effects require to use stronger and more
localized perturbations, where a LL description is not applicable.
In addition, the effect of boundaries, where the local density
approximation breaks down, are of crucial importance. For a
quantitative description of spin-charge separation in 1D cold
Fermi gases, it is thus necessary to use a microscopic description
like the Hubbard model and properly treat the inhomogeneous case
with realistic system sizes. Due to the short range nature of the
interactions between cold atoms, the Hubbard  model is indeed a
perfect description of a situation, in which there is an
additional optical lattice along the weakly confined axial
direction (for bosons, the corresponding setup has already been
realized, see \cite{Stoeferle2003, ParedesBloch2004}). It is an
essential new feature of cold atoms in optical lattices that
parameters can be changed dynamically and the resulting time
evolution can be studied. This gives direct access to the
real-time dynamics of strongly correlated systems, a subject
hardly studied so far. In this context, a successful method is the
recently developed adaptive time-dependent density matrix
renormalization group (adaptive t-DMRG) \cite{Dale04} which is an
efficient implementation of Vidal's TEBD algorithm \cite{Vida03}
in the DMRG framework \cite{Whit92}. It has previously been
applied to study density perturbations in bosonic 1D condensates
over a large range of interaction strengths \cite{Koll04}. The
real-time dynamics of interacting spinful Fermionic systems is
much harder and has not been studied except for very small systems
\cite{Jagl93}. In our present work, we present numerical results
of the real-time dynamics of a 1D Hubbard model for realistic
sizes of up to 128 sites. Our main results are:
\\(i) real-time calculation showing spin-charge separation explicitly
in systems of %realistic and
experimentally accessible size % \cite{Jagl93}.
\\(ii) the demonstration that spin-charge separation survives far outside the
low-energy LL regime %\cite{Zacher}
\\(iii) a quantitative calculation for the effect of spin-charge separation at the
boundary between a liquid and a Mott-insulating (MI) phase which allows
to observe the phenomenon in cold gases without the problems
arising from the different densities in an array of parallel atomic wires
and to distinguish experimentally between a Mott- and a band insulator.
% {\bf vielleicht so etwas wie:
%We also assured by calcuation that the
%  time-evolution of a single-particle excitation on a certain site $j$,
%  created by the application of $c^\dagger_j$, decays rapidly into a charge
%  and a spin density wave with different velocities.}

Our starting point is the standard Hubbard model
\begin{eqnarray}
\label{eq:fh}
H = &&-J \sum_{j,\sigma}\left( c^\dagger_{j+1,\sigma}c_{j,\sigma} + h.c.\right) +
U \sum_j n_{j,\uparrow}n_{j,\downarrow}\nonumber \\&&+ \sum_{j,\sigma} \varepsilon_{j,\sigma} \hat{n}_{j,\sigma}
\end{eqnarray}
for Fermions in 1D. Its parameters are the hopping matrix element
$J$, the on-site repulsion $U>0$ between Fermions of opposite spin
$\sigma=\uparrow, \downarrow$ at sites $j=1,\ldots, L$ and a
spin-dependent local on-site energy $\varepsilon_{j,\sigma}$,
describing both a possible smooth harmonic confinement and
time-dependent local potentials.%%% which allow to perturb the system.
One introduces a 'charge' density $n_c=n_\uparrow
+n_\downarrow$ and a 'spin' density $n_s=n_\uparrow
-n_\downarrow$; in a realization with cold gases, the spin degrees
of freedom are represented by two different hyperfine levels, and
'charge' density is particle density. Similar to bosons in an
optical lattice \cite{Jaksch98}, the ratio $u=U/J$ %%%between the
%%%on-site repulsion $U$ and the hopping $J$
can easily be changed experimentally by varying the depth $V_0$ of
the optical lattice. We use units where both $J$ and $\hbar$ are
equal to one; thus time is measured in units of $\hbar/J$. In the
numerical calculations below, we study the dynamics of the Hubbard
model using different initial density perturbations and the
excitations resulting from adding a single particle at a given
lattice site, which is expected to display the same physics as
contained in single particle spectral functions.% (see below).
Experimentally, the density perturbations may be generated by a
blue- or red-detuned laser beam tightly focused perpendicular to
an array of atomic wires, which generates locally repulsive or
attractive potentials for the atoms in the wires. In practice, the
perturbations due to an external laser field are quite strong,
typically of the order of the recoil energy $E_r$ and thus clearly
require a nonperturbative treatment. In our calculations the
length of the chains was chosen up to $L=128$ sites, keeping of
the order of several hundred DMRG states. DMRG error analysis
reveals that all density distributions shown here are exact for
all practical purposes, with controlled errors of less than
$O(10^{-3})$ \cite{gob04}.

\begin{figure}
\begin{center}
         {\epsfig{figure=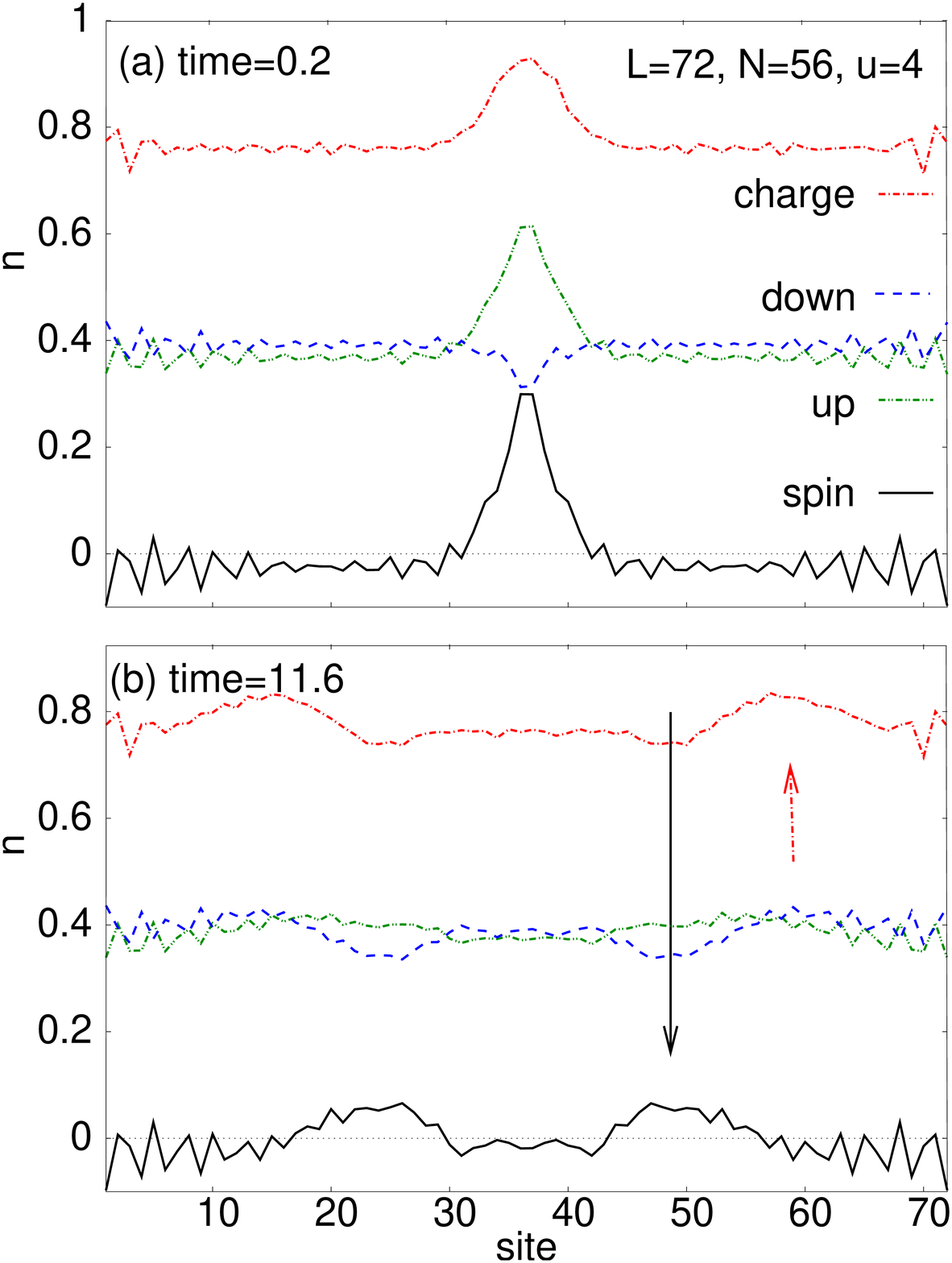,width=0.6\linewidth}}
\end{center}
\caption{Snapshots of the evolution of the density distribution are shown at
  different times. At $t=0.2$, a wave packet is present in the center of
  the system in both the spin and the charge density. Each of these splits up into two packets which move with the same velocity in
  opposite directions. The velocity of the charge wave and the spin wave
  are different.%%% $U/J=4$,  background density is $n_0=0.78$.
  }
\label{fig:snapshot}
\end{figure}

We start with a homogeneous system which is perturbed by a
potential $\varepsilon_{j,\uparrow}$ localized at the chain center which
couples only to the $\uparrow$-Fermions, i.e.
\begin{equation}
\label{eq:gauss}
\varepsilon_{j,\uparrow}(t)\propto  \exp{ \{ -[j-(L-1)/2]^2/8 \} }
\, \theta (- t) \;
 \end{equation}
The potential is assumed to have been switched on slowly enough
for equilibration, and is then switched off suddenly at time $t =
0$. In Fig.~ \ref{fig:snapshot} (a) the density distribution of
the
state at $t=0.2$ is shown.%%% as obtained by DMRG.
 The external
potential (\ref{eq:gauss}) generates a dominant perturbation in
the $\uparrow$-Fermion distribution by direct coupling and,
indirectly, a smaller perturbation in the $\downarrow$-density due
to the repulsive interaction between the different spin species.
The wave packets in $\uparrow$ and $\downarrow$-density hence
perform a complicated time evolution (Fig.~ \ref{fig:snapshot}).
In contrast, the perturbations in the spin and charge density
split into two wave packets each moving outwards. Their respective
velocities are %found to be
different as indicated by the arrows in Fig.~ \ref{fig:snapshot}
(b), separating spin and charge.

\begin{figure}
\begin{center}
         {\epsfig{figure=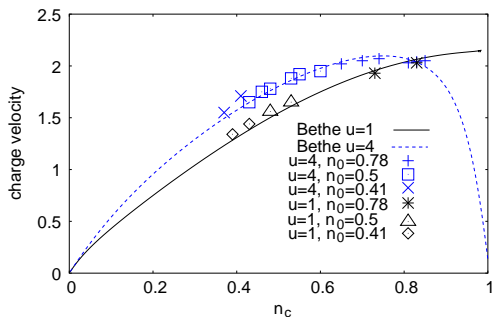,width=0.75\linewidth}}
\end{center}
\caption{Exact results for the charge velocity obtained by
  the Bethe ansatz are (lines) compared to the
  numerical results of the adaptive t-DMRG. The numerical results
  correspond to different heights of the perturbations at various
  charge background densities $n_0$. $n_c$ is the charge density at
  the maximum/minimum of the charge density perturbation. The
  uncertainties are of the order
  of the size of the symbols and stem mainly from the determination of
  the velocity.
}
\label{fig:bethe}
\end{figure}

In the limit of an infinitesimal perturbation much broader than
the average interparticle spacing, both spin and charge velocities
are known analytically from the Bethe ansatz
\cite{LiebWu1968}. To compare our numerical findings
to the exact charge velocity, we create pure charge density perturbations, by
applying the potential of Eq. (\ref{eq:gauss}) to both species,
i.e. $\varepsilon_{j,\uparrow}= \varepsilon _{j,\downarrow}$,
and calculate their time-evolution after switching off the potential.
The charge velocity is determined from the propagation of the maximum (minimum) of
the charge density perturbation for bright (amplitude $\eta_c>0$) and grey ($\eta_c<0$)
perturbations, respectively. In Fig. \ref{fig:bethe} the charge velocities for various
background densities $n_0$ and perturbation amplitudes $\eta_c$ are
shown. We find good agreement, if we plot the charge velocity versus the
charge density at the maximum (minimum), i.e. $n_c=n_0+\eta_c$.
%%%The velocity of the maximum (minimum) of the
%%%wave packet is therefore determined by the value of the charge density
%%%at the maximum (minimum), not the background density.
This stays true even for strong perturbations $\eta_c\approx \pm
0.1$ which corresponds to 20\% of the charge density. The charge
velocity is thus robust against separate changes of the background
density $n_0$ and the height of the perturbation $\eta_c$. By
contrast, the velocity of a spin perturbation varies strongly with
its height. A possible reason for this may be the nonlinear
$\cos{\phi}$-contribution in the LL description of the spin
density field, which is only marginally irrelevant, giving rise to
a nonanalytic contribution to the spin-density response
\cite{Giam04}. Nevertheless,
 our numerical results show that for
decreasing height of the spin perturbation, its velocity approaches
the value for the spin velocity obtained by the Bethe ansatz.

\begin{figure}
\begin{center}
         {\epsfig{figure=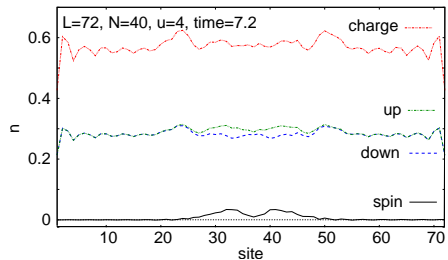,width=0.67\linewidth}}
\end{center}
\caption{Snapshot of the time-evolution of the charge and spin
  densities of a single particle
  excitation created at time $t=0$ at site $j=37$ is shown for $t=7.2$.
}
\label{fig:singleexc}
\end{figure}

In order to compare the behaviour of the density
perturbations with that of a single particle excitation, the time evolution of
the system with one additional particle added at time $t=0$ on site $j$ to the ground state, is
calculated numerically.  In Fig. \ref{fig:singleexc} a snapshot of the resulting
evolution of the densities
is shown for time $t=7.2$. Remarkably, as in the case of the density perturbation
%after a short time
 separate wave packets in spin and charge can be seen.
This demonstrates the phenomenon of spin-charge separation directly
in a single particle excitation, in close analogy to the situation
of an inverse photo-emission experiment for the single particle spectral functions
\cite{Sego99}

\begin{figure}
\begin{center}
         {\psfig{figure=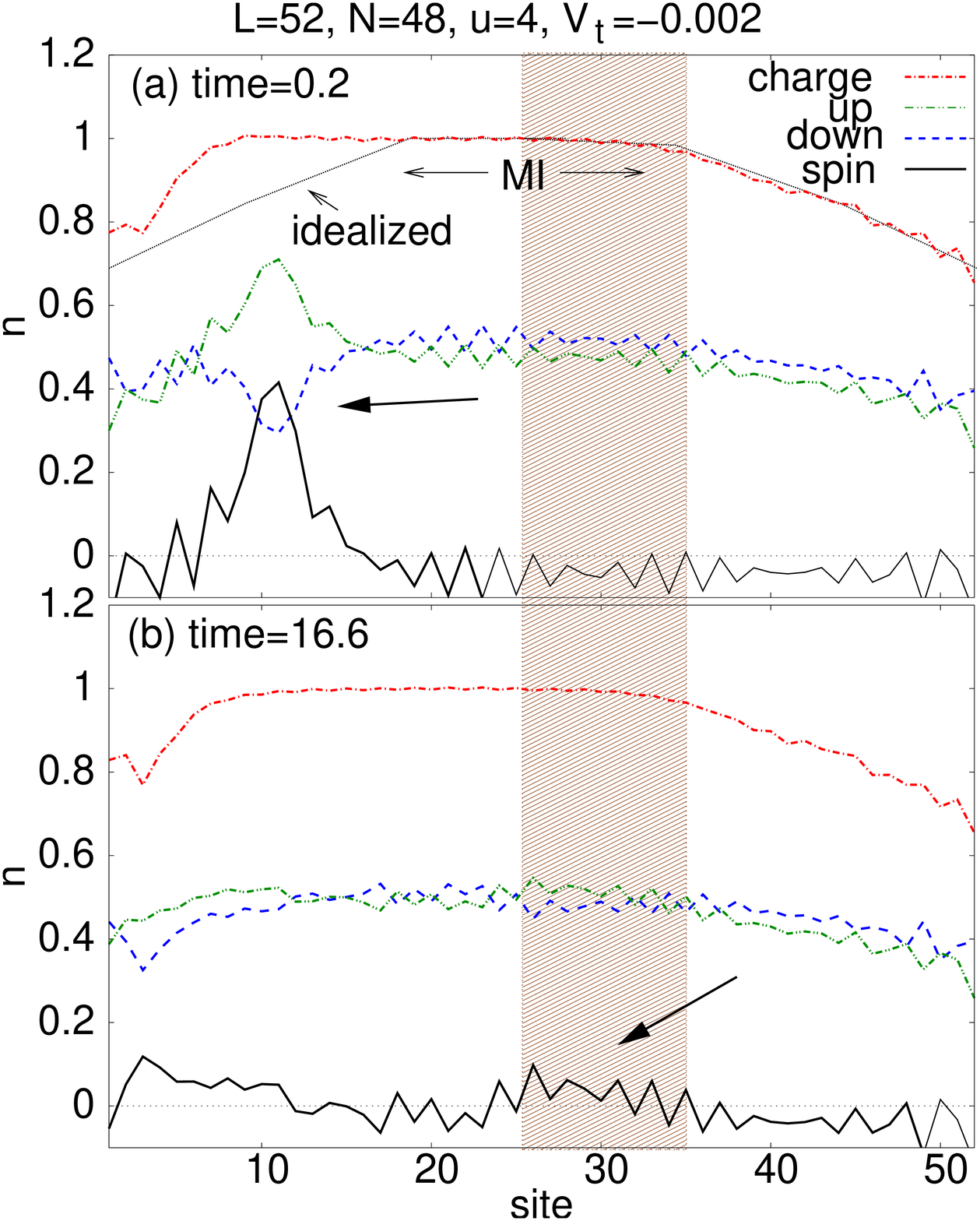,width=0.7\linewidth}}
\end{center}
\caption{Time-evolution of charge and spin density perturbations
  in the presence of a parabolic trapping potential
  $\varepsilon_{\sigma,j}= -V_t a^2(j-L/2+0.5)^2 E_r$. MI
  marks the approximate MI region in the absence of the
  perturbation. The line denoted by idealized is a sketch of the
  charge density distribution without the perturbation. The presence of the perturbation enlarges the region
  in which the charge density is locked to $n_c=1$. The arrows show the
  approximate place of the spin perturbation, and the shaded region marks the
  region over which the densities are averaged (cf. Fig. \ref{fig:trap}).}
\label{fig:snaptrap}
\end{figure}

\begin{figure}
  \begin{center}
    {\epsfig{figure=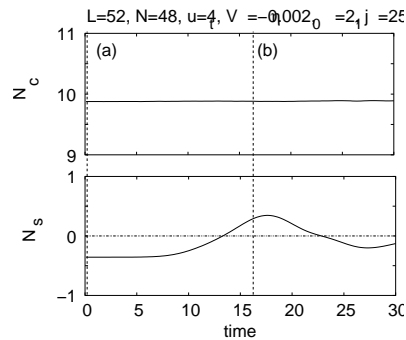,width=0.6\linewidth}}
  \end{center}
  \caption{Time evolution of the charge and spin density summed over the sites $j_0= 25$ to
    $j_1=35$. Vertical
    lines correspond to the times of the snapshots in Fig. \ref{fig:snaptrap}}
  \label{fig:trap}
\end{figure}

In a specific experiment with arrays of parallel atomic wires, it
is necessary to take into account that there is an additional
harmonic trapping potential. Moreover, individual wires have
slightly different fillings, which leads to an inhomogeneous
broadening due to the resulting difference in velocities. In order
to observe an unambiguous signal of spin-charge separation in such
a situation, we suggest an experimental setup, which relies on the
coexistence of a MI state and a liquid state in spatially
separated regions of the
  parabolically confined system \cite{RigolMuramatsu2004}. The idea is
  to use the very different behaviour of the charge and spin degrees
  of freedom in the MI phase. In this phase the charge excitation spectrum has a gap, whereas
the spin dispersion is still linear for small momenta, and the spin velocity is
finite.
By contrast, in the liquid phase both excitation spectra are linear for small
momenta. To exploit this, assume the system of
one-dimensional wires is prepared in such a way, that  a
MI region is present in the center, where the charge
density is locked at half-filling, $n_c=1$. At the boundary of this
MI region liquid regions appear.
A localized potential in the liquid region will then create spin and charge density
waves. Calculated snapshots of the time evolution in such a situation are
shown in Fig. \ref{fig:snaptrap}. Evidently,
the spin density wave propagates into the
MI region whereas the charge density perturbation is almost completely
reflected due to the charge gap in the MI. The presence of spin density
oscillations which are due to the antiferromagnetic coupling induced by the
interaction obscures the exact evolution of the spin perturbation.
However by averaging  over several lattice sites - as is always necessary
in an experiment - the effect of spin-charge separation is clearly visible.
In Fig. \ref{fig:trap} examples for the evolution of
the sum of the charge and the spin number of particles between site 25
and 35, $N_c$ and $N_s$ respectively, are shown. It is clearly seen that the sum of
the charge occupation does not change,
whereas the spin occupation shows the moving wave packet. The average spin velocity
can be determined from Fig. \ref{fig:trap} if the distance between
the localized potential which generates the perturbation and the
center of the region over which the density is measured is known.
Here, the spin velocity is found to be $v_s\approx 1.1 J/ \hbar$ which agrees
nicely within the expected accuracy with the value of $v_s(n_c=1)=1.2
J/\hbar$ of the Bethe ansatz.
The very different propagation behaviour of
charge and spin can as well be used experimentally
to distinguish between a MI and a band insulator:
In a band insulator not only the velocity of the
charge, but as well of the spin would vanish, whereas, as used above, in the
MI the spin velocity stays finite.

In order to quantify the requirements for an experimental
observation of spin-charge separation in cold Fermi gases, we
finally discuss typical parameters which need to be achieved in a
setup with an array of atomic wires \cite{Mori05}. Such an array
consists of several thousand parallel wires with typically less
than 100 $^{40}\textrm{K}$ atoms each. In addition to the smooth
axial confinement potential with frequency
$\omega_z\approx\varepsilon_F/N\approx 2\pi\cdot 275 \textrm{Hz}$
(corresponding to $V_t \approx -0.0035$), realization of a 1D
Hubbard model requires adding a strong periodic potential along
the tubes. For $^{40}\textrm{K}$ and  a standard lattice constant
$a=413 \textrm{nm}$ the recoil energy is $E_r\approx 7
\textrm{kHz}$. An optical lattice of height $V_0=15 E_r$ then
gives an on-site repulsion $U\approx 0.17 E_r$, where we have used
a standard value for the  s-wave scattering length $a_s \sim 174
a_0$ for the $F=9/2$ $m_f=-9/2$ and $m_f=-7/2$ states
\cite{RegalJin2003}. The resulting dimensionless interaction
$u\approx 22$ then leads to a central MI region with a typical
size of around 20 sites. With this parameters, the time in which
the spin wave travels 20 sites is of the order of a few ms. The
creation of state selective potentials for two different hyperfine
states may be done by using laser light whose frequency falls
between the respective transitions. This might be difficult for
the $F=9/2$ $m_f=-9/2$ and $m_f=-7/2$ levels, but should be
possible  - for instance- using the $F=9/2$ and $F=7/2$ levels.
The $1/e^2$-radius of the potential (Eq. \ref{eq:gauss}) is taken
to be four lattice sites, which could be realized approximately by
a laser of an $1/e^2$-radius of $2.1 \mu m$ or less. Finally, to
ensure that finite temperature does not destroy the Mott
insulating behaviour by thermal activation, the energy scale
$k_BT$ should be smaller than the Mott energy gap. Already the
very first experiment of 1D fermions in an optical lattice is very
close to matching those conditions. Improvements can be reached
reducing the axial confinement frequency.

To conclude we have performed numerical simulations of the time-evolution of charge
and spin density perturbations in the 1D Hubbard model.  We clearly observe
the separation of spin and charge as a generic feature of 1D fermions, far beyond the low-energy
regime where a Luttinger liquid description applies. In addition, an  experiment is suggested
which exhibits the separation of the two modes from the perfect reflection of
density excitations at the boundary to a Mott insulating state

We gratefully acknowledge very fruitful discussions
with Th. Giamarchi and with the experimental group of T. Esslinger, in particular
H. Moritz.

\end{document}